\documentclass[conference]{IEEEtran}

\usepackage[cmex10]{amsmath}
\usepackage[pdftex]{graphicx}
\usepackage{array}
\usepackage{mdwmath}
\usepackage{mdwtab}
\usepackage{amssymb}
\usepackage{float}
\usepackage{subfig}

\newtheorem{theorem}{Theorem}

\newtheorem{proposition}[theorem]{Proposition}
\newcommand{\qed}{\hfill$\square$}

\DeclareMathAlphabet{\bm}{OML}{cmm}{b}{it}

\newcommand{\bol}[1]{\mathbf{#1}}
\newcommand{\rom}[1]{\mathrm{#1}}
\newcommand{\san}[1]{\mathsf{#1}}

\usepackage[ruled,vlined]{algorithm2e}
\usepackage{algorithmic}
\newenvironment{protocol}[1][htb]
  {%
   \begin{algorithm}[#1]%
  }{\end{algorithm}}

\begin{document}
%
\title{An Improved Lower Bound on Oblivious Transfer Capacity Using Polarization and Interaction}



%
\author{\IEEEauthorblockN{So Suda and Shun Watanabe}
\IEEEauthorblockA{Department of Computer and Information Sciences, 
Tokyo University of Agriculture and Technology, Japan, \\
E-mail:s238948y@st.go.tuat.ac.jp; shunwata@cc.tuat.ac.jp}}


\maketitle

\begin{abstract}
We consider the oblivious transfer (OT) capacities of noisy channels against the passive adversary;
this problem has not been solved even for the binary symmetric channel (BSC).
In the literature, the general construction of OT has been known only for generalized erasure channels (GECs);
for the BSC, we convert the channel to the binary symmetric erasure channel (BSEC), which is a special instance 
of the GEC, via alphabet extension and erasure emulation. In a previous paper by the authors, we derived an improved
lower bound on the OT capacity of BSC by proposing a method to recursively emulate BSEC via interactive communication.
In this paper, we introduce two new ideas of OT construction: (i) via ``polarization" and interactive communication,
we recursively emulate GECs that are not necessarily a BSEC; (ii) in addition to the GEC emulation part, 
we also utilize interactive communication in the key agreement part of OT protocol. By these methods,
we derive lower bounds on the OT capacity of BSC that are superior to the previous one for a certain range of
crossover probabilities of the BSC. Via our new lower bound, we show that,
at the crossover probability being zero, the slope of tangent of the OT capacity
is unbounded.
\end{abstract}


%
\IEEEpeerreviewmaketitle

\section{Introduction} \label{section:introduction}

One of the most important problems in modern cryptography is the secure computation introduced by Yao \cite{Yao82}.
In this problem, the parties seek to compute a function in such a manner that the parties do not learn any additional information
about the inputs of other parties more than the output of the function value itself.    
For two-party secure computation, 
non-trivial functions are not securely computable from scratch \cite{Bea89b, Kus92} (see also \cite{NTW15}).
In order to securely compute non-trivial functions, we usually assume the availability of
a primitive termed the oblivious transfer (OT) \cite{EveGolLem85}. If the OT is available, then
any functions are securely computable \cite{GolVai88, Kilian88}.

In late 80's, it was shown that information-theoretically secure oblivious transfer can be implemented
from noisy channels between the parties \cite{CreKil88}. 
Since then, characterizations of resources to realize secure computation is regarded as an important problem, 
and it has been actively studied \cite{MajPraRos:09, MajPraRos12, MajPraRos13, KraQua:11, KraMajPraSah:14}.
In addition to the feasibility of secure computation, it is also important 
to characterize the efficiency of realizing secure computation from a given resource.
Toward that direction, Nascimento and Winter 
introduced the concept of the OT capacity \cite{NasWin08}, which is defined as the length
of the string oblivious transfer that can be implemented per channel use.
Later, more general upper and lower bounds on the OT capacity were further studied by
Ahlswede and Csisz\'ar in \cite{AhlCsi13}. Based on the tension region idea in \cite{PraPra14},
Rao and Prabhakaran derived the state-of-the-art upper bound on the OT capacity in \cite{RaoPra14}. 

In this paper, we revisit the OT capacity of noisy channels against the passive adversary.
Despite an effort in the literature, the OT capacity is not known even for the most basic channel,
the binary symmetric channel (BSC).

The general construction of oblivious transfer has been known only for generalized
erasure channels (GECs).
The binary symmetric channel (BSC) is a typical example that is not a GEC.
Thus, we first convert the BSC to the binary symmetric erasure channel (BSEC), which is a special instance of the GEC,
via alphabet extension and erasure emulation, and then apply the general construction for GEC \cite{AhlCsi13}. 
In our previous work \cite{SudWatYam:24}, we proposed a method to recursively emulate BSEC via interactive
communication, and we derived an improved lower bound on the OT capacity that is superior to the one 
derived by a naive emulation of BSEC in \cite{AhlCsi13}. 

In this paper, we introduce two new methods of OT construction.
The idea of the first method is to depart from BSEC emulation and to emulate more general GECs.
When we emulate BSEC, we only use inputs $(0,0)$ or $(1,1)$ for two invocations of BSCs; this sacrifice
of degree of freedom of inputs  enables us to regard the received signals $\{(0,1),(1,0)\}$ as erasure since
a posteriori probability of $(0,0)$ and $(1,1)$ are the same. A difficulty of emulating GECs other than BSEC
is how to find an appropriate set of received signals that plays a role of erasure while controlling
a sacrifice of degree of freedom of inputs. We propose a method to emulate GECs that is inspired by
the polarization of polar code \cite{arikan:09}. By this method, we derive an improved lower bound on the OT capacity
that is superior to the one in \cite{SudWatYam:24} for a certain range of the crossover probability.
Particularly, as the size of polarization block increases, we show that, at the crossover probability being zero,
the slope of tangent of the lower bound can be arbitrarily large; this matches the behavior of known upper bound \cite{AhlCsi13}. 

The idea of the second method is to use the interactive key agreement \cite{watanabe:07,gohari:10}; see also \cite[Sec.~10.5]{TyaWat:book}.
Even though interactive communication has been used in the GEC emulation part of existing OT protocols, the key
agreement part of existing OT protocols is one-way. In the context of key agreement, interactive protocols are known to
be superior to one-way protocols in certain situations. However, in the key agreement part of OT construction, 
the receiver plays roles of both the legitimate party and the adversary in parallel, and the receiver must behave the same
in both the roles so that the receiver's input of OT is not leaked to the sender. We overcome this subtlety 
by proposing a protocol in which the adversary role of the receiver sends a dummy message to imitate
the legitimate party role of the receiver.
By this method, we further improve the lower bound obtained by the first method above. 



\section{Problem Formulation} \label{section:problem}

We mostly follow the notations from \cite{csiszar-korner:11, TyaWat:book}.
Random variables and random vectors are denoted by capital letters such as $X$ or $X^n$; 
realizations are denoted by lowercase letters such as $x$ or $x^n$; 
ranges are denoted by corresponding calligraphic letters such as ${\cal X}$ or ${\cal X}^n$;
for an integer $n$, the index set is denoted as $[n] = \{1,\ldots,n\}$; for a subset ${\cal I} \subset [n]$
and a random vector $X^n = (X_1,\ldots,X_n)$ of length $n$, the random vector $X_{{\cal I}}$ is a collection
of $X_i$s such that $i \in {\cal I}$. The entropy and conditional entropies are denoted as $H(X)$ and $H(X|Y)$;
the binary entropy function is denoted as $h(q)= q\log \frac{1}{q}+(1-q)\log\frac{1}{(1-q)}$ for $0 \le q \le 1$.

Let $W$ be a channel from a finite input alphabet ${\cal X}$ to a finite output alphabet ${\cal Y}$.
By using the channel $n$ times, the parties, the sender ${\cal P}_1$ and the receiver ${\cal P}_2$,
shall implement the string OT of length $l$. More specifically, the sender generates two
uniform random strings $K_0$ and $K_1$ on $\{0,1\}^l$, and the receiver generates a uniform bit $B$ on $\{0,1\}$,
as inputs to OT protocol. In the protocol, in addition to communication over the noisy channel $W^n$,
the parties are allowed to communicate over the noiseless channel, possibly multiple rounds.\footnote{In general,
the communication over the noiseless channel may occur between invocations of the noisy channels; for a more precise 
description of OT protocols over noisy channels, see \cite{AhlCsi13}.}
Let $\Pi$ be the exchanged messages over the noiseless channel, and let $X^n$ and $Y^n$ be the input
and the output of the noisy channel $W^n$. At the end of the protocol, ${\cal P}_2$ computes an estimate
$\hat{K} = \hat{K}(Y^n,\Pi, B)$ of $K_B$. 

For $0 \le \varepsilon,\delta_1,\delta_2 <1$,
we define that a protocol realizes $(\varepsilon,\delta_1,\delta_2)$-secure OT of length $l$ (for passive adversary) if
the following three requirements are satisfied:
\begin{align}
\Pr(\hat{K} \neq K_B) &\le \varepsilon, \label{eq:correct} \\
d_{\mathtt{var}}(P_{K_{\overline{B}} Y^n \Pi B}, P_{K_{\overline{B}}} \times P_{Y^n \Pi B}) &\le \delta_1, \label{eq:security-1} \\
d_{\mathtt{var}}(P_{BK_0K_1X^n\Pi}, P_B \times P_{K_0 K_1 X^n \Pi}) &\le \delta_2, \label{eq:security-2}
\end{align}
where $d_{\mathtt{var}}(P,Q)=\frac{1}{2}\sum_a|P(a)-Q(a)|$ is the variational distance. 
The requirement \eqref{eq:correct} is referred to as $\varepsilon$-correctness; the requirement \eqref{eq:security-1}
is the security for ${\cal P}_1$ in the sense that $K_{\overline{B}}$ is concealed from ${\cal P}_2$ observing $(Y^n,\Pi, B)$,
where $\overline{B}=B\oplus 1$;
and the requirement \eqref{eq:security-2} is the security for ${\cal P}_2$ in the sense that $B$ is concealed from 
${\cal P}_1$ observing $(K_0,K_1,X^n,\Pi)$.

A rate $R$ is defined to be achievable if, for every $0 \le \varepsilon,\delta_1,\delta_2 <1$ and sufficiently large $n$,
there exists an $(\varepsilon,\delta_1,\delta_2)$-secure OT protocol of length $l$ satisfying $\frac{l}{n}\ge R$.
Then, the OT capacity $C_{\mathtt{OT}}(W)$ is defined as the supremum of achievable rates.

\section{Review of Standard Protocol for GEC} \label{section:GEC}

\begin{protocol}[h] 
\caption{Standard Protocol for GEC} \label{protocol-GEC}
\begin{algorithmic}[1]
\STATE \label{protocol-GEC-step1}
${\cal P}_1$ sends $X^n$ over the GEC $W^n$, and ${\cal P}_2$ receives $Y^n$.

\STATE \label{protocol-GEC-step2}
For each $i\in[n]$, ${\cal P}_2$ generates $V_i \in \{0,1,2\}$ as follows:
\begin{itemize}
\item If $Y_i \in {\cal Y}_1$, then set $V_i=1$;

\item 
If $Y_i \in {\cal Y}_0$, then set $V_i=0$ with probability $\frac{p}{1-p}$ and set $V_i=2$ with probability $\frac{1-2p}{1-p}$.
\end{itemize}

\STATE \label{protocol-GEC-step3}
Then, ${\cal P}_2$ sets $\tilde{{\cal I}}_b = \{ i \in [n]: V_i = b\}$ for $b=0,1$. If 
$|\tilde{{\cal I}}_0| < m$ or $|\tilde{{\cal I}}_1| < m$, then abort the protocol. 

\STATE \label{protocol-GEC-step4}
${\cal P}_2$ sets
${\cal I}_B$ as the first $m$ indices from $\tilde{{\cal I}}_0$ and ${\cal I}_{\overline{B}}$ as the first $m$ indices from $\tilde{{\cal I}}_1$,
and sends $\Pi_1 = ({\cal I}_0, {\cal I}_1)$ to ${\cal P}_1$.

\STATE \label{protocol-GEC-step5}
${\cal P}_1$ randomly picks functions $F:{\cal X}^m \to \{0,1\}^l$ and $G:{\cal X}^m \to \{0,1\}^\kappa$
from universal hash families, computes $S_b = F(X_{{\cal I}_b})$ and $C_b = G(X_{{\cal I}_b})$ for $b=0,1$,
and sends $\Pi_2 = (\Pi_{2,0}, \Pi_{2,1}, \Pi_{2,2})$, where 
\begin{align*}
\Pi_{2,b} = K_b \oplus S_b \mbox{ for } b=0,1, \mbox{ and } \Pi_{2,2}=(C_0,C_1).
\end{align*}

\STATE \label{protocol-GEC-step6}
${\cal P}_2$ reproduces $\hat{X}_{{\cal I}_B}$ from $C_B$ and $Y_{{\cal I}_B}$, and computes $\hat{K}= \Pi_{2,B} \oplus F(\hat{X}_{{\cal I}_B})$.
\end{algorithmic}
\end{protocol}

In this section, we review the standard OT protocol for generalized erasure channel (GEC) 
introduced in \cite{AhlCsi13}. 
A GEC consists of a partition ${\cal Y}={\cal Y}_0 \cup {\cal Y}_1$ of output alphabet and erasure probability $0 < p \le \frac{1}{2}$;\footnote{Even though
the GEC can be defined for $\frac{1}{2} < p \le 1$, we focus on $0 < p \le \frac{1}{2}$ in this paper.}
 the transition
matrix is given by
\begin{align*}
W(y|x) = \left\{
\begin{array}{ll}
(1-p) W_0(y|x) & \mbox{if } y \in {\cal Y}_0 \\
p W_1(y|x) & \mbox{if } y \in {\cal Y}_1
\end{array}
\right.
\end{align*}
for two channels $W_0:{\cal X}\to {\cal Y}_0$ and $W_1:{\cal X}\to {\cal Y}_1$.
For instance, the binary symmetric erasure channel is a GEC such that ${\cal Y}_1 = \{\mathsf{e}\}$ for the erasure symbol $\mathsf{e}$
and $W_0$ is the binary symmetric channel. Even though ${\cal Y}_1$ needs not be $\{\mathsf{e}\}$ in general, we call ${\cal Y}_1$ as
the {\em erasure set} and ${\cal Y}_0$ as the {\em non-erasure set} throughout the paper.

Before describing the OT protocol, let us review two building blocks known as 
the information reconciliation and the privacy amplification (e.g.~see \cite[Chapters 6 and 7]{TyaWat:book} for more detail):
\paragraph*{Information Reconciliation} We consider the situation such that the sender and the receiver observe correlated i.i.d. sources $X^m$ and $Y^m$;
the sender transmits a message $\Pi$ to the receiver, and the receiver reproduces an estimate of $X^m$ by using the transmitted message and the side-information
$Y^m$. This problem is known as the source coding with side information (Slepian-Wolf coding with full side information), and it is known that
the receiver can reproduce $X^m$ with small error probability if, for some margin $\Delta>0$,  
the sender transmits a message of length $\kappa= \lceil m(H(X|Y)+\Delta)\rceil$ that is created by the universal hash family.

\paragraph*{Privacy Amplification} It is a procedure to distill a secret key from a randomness that is partially known to the adversary.
In the OT construction, we consider the situation such that ${\cal P}_1$ and ${\cal P}_2$ (who plays the role of the adversary)
observe correlated i.i.d. sources $X^m$ and $Z^m$; furthermore, the adversary may observe additional message $\Pi$ (obtained 
during the information reconciliation). Then, for a randomly chosen function $F$ from universal hash family, the key $K=F(X^m)$ 
is almost uniform and independent of the adversary's observation $(Z^m,\Pi)$ provided that the length of the generated key 
is $l = \lfloor m(H(X|Z)- \Delta)\rfloor - \kappa$ for some margin $\Delta>0$, where $\kappa$ is the length of additional message $\Pi$.

\paragraph*{Standard Protocol} The high-level flow of the standard protocol is as follows. 
First, ${\cal P}_1$ sends a sequence of symbols $X^n$ over the GEC 
and ${\cal P}_2$ receives $Y^n$. Then, ${\cal P}_2$ picks subsets ${\cal I}_0, {\cal I}_1 \subset [n]$ so that ${\cal I}_B$ consists of indices
such that ${\cal P}_2$'s observation is included in the non-erasure set ${\cal Y}_0$ 
and ${\cal I}_{\overline{B}}$ consists of indices such that 
${\cal P}_2$'s observation is included in the erasure set ${\cal Y}_1$. Since the event $Y_i \in {\cal Y}_1$ occurs obliviously to ${\cal P}_1$,
revealing the index sets $({\cal I}_0,{\cal I}_1)$ to ${\cal P}_1$ does not leak any information about $B$. 
On the other hand, by using randomly chosen functions from universal hash families, ${\cal P}_1$ generates a pair of secret keys
$(S_0,S_1)$, and sends $(K_0, K_1)$ to ${\cal P}_2$ by encrypting with the generated keys so that ${\cal P}_2$ can only decrypt $K_B$. 
Formally, the standard protocol is described in Protocol \ref{protocol-GEC}. 
Since $0< p \le \frac{1}{2}$, part of non-erasure indices are discarded.\footnote{For $\frac{1}{2} < p <1$, 
we need to modify the discarding rule in Step \ref{protocol-GEC-step2}.}
In the protocol, the random variable $V_i$ describes that the index $i$
is erasure if $V_i=1$; it is non-erasure and not discarded if $V_i=0$; and it is non-erasure but is discarded if $V_i=2$.

Now, we outline the security and the performance of Protocol \ref{protocol-GEC}.
First, to verify that ${\cal P}_2$'s message $\Pi_1=({\cal I}_0,{\cal I}_1)$ does not leak any information about $B$ to ${\cal P}_1$,
let us introduce $\tilde{V}_i = V_i \oplus B$ if $V_i \in \{0,1\}$ and $\tilde{V}_i = V_i$ if $V_i=2$.
Then, note that ${\cal I}_0$ and ${\cal I}_1$ are the first $m$ indices of $\{ i : \tilde{V}_i=0\}$
and $\{ i : \tilde{V}_i=1\}$, i.e., $({\cal I}_0,{\cal I}_1)$ is a function of $\tilde{V}^n$.
Thus, it suffices to show that $I(X^n, \tilde{V}^n \wedge B)=0$. Since $I(X^n \wedge B) = 0$,
we will verify $I(\tilde{V}^n \wedge B|X^n)=0$.
Since $0 < p \le \frac{1}{2}$, note that\footnote{See \cite[Sec.~13.3.1]{TyaWat:book} for more detail.}
\begin{align*}
\Pr( V_i = 0 | X_i = x) = \Pr(V_i = 1 | X_i = x) = p
\end{align*}
for every $i \in [n]$, which implies
\begin{align*}
P_{\tilde{V}_i | X_i B}(v|x,0) &= P_{V_i|X_i}(v|x) \\
&= P_{V_i|X_i}(v \oplus 1|x) 
= P_{\tilde{V}_i|X_i B}(v|x,1)
\end{align*}
for $v \in \{0,1\}$. Also, we have $P_{\tilde{V}_i|X_i B}(2|x,b) = P_{V_i|X_i}(2|x)$. Thus, 
we have $I(\tilde{V}^n \wedge B|X^n) = 0$.

For a small margin $\Delta >0$, if we set $m = \lfloor n(p-\Delta) \rfloor$, then
the protocol is not aborted in Step \ref{protocol-GEC-step3} with high probability.
Furthermore, since the channel between $X_i$ and $Y_i$ conditioned on $V_i=0$ is
$P_{Y_i|X_i V_i}(y|x,0) = W_0(y|x)$,
if we set $\kappa = \lceil m(H(\tilde{X}|\tilde{Y}_0) + \Delta) \rceil$ for output $\tilde{Y}_0$ of channel $W_0$ with input $\tilde{X}$, 
then, by the result on the information reconciliation,
${\cal P}_2$ can reproduce $X_{{\cal I}_B}$ with small error probability in Step \ref{protocol-GEC-step6}.
Finally, since the channel between $X_i$ and $Y_i$ conditioned on $V_i=1$ is
$P_{Y_i|X_i V_i}(y|x,0) = W_1(y|x)$,
if we set $l = \lfloor m(H(\tilde{X}|\tilde{Y}_1)-\Delta) \rfloor - \kappa$ for output $\tilde{Y}_1$ of channel $W_1$ with input $\tilde{X}$, 
then, by the result on the privacy amplification,
the key $S_{\overline{B}}$ is almost uniform and independent of ${\cal P}_2$'s observation;
since $K_{\overline{B}}$ is encrypted by the one-time pad with key $S_{\overline{B}}$, $K_{\overline{B}}$ is not leaked to ${\cal P}_2$.
Consequently, 
by taking $\Delta>0$ sufficiently small and $n$ sufficiently large, Protocol \ref{protocol-GEC} realizes
$(\varepsilon,\delta_1,\delta_2)$-secure OT of length roughly $np(H(\tilde{X}|\tilde{Y}_1)-H(\tilde{X}|\tilde{Y}_0))$, i.e., we have
\begin{align} \label{eq:OT-capacity-GEC}
C_{\mathtt{OT}}(W) \ge p (H(\tilde{X}|\tilde{Y}_1)-H(\tilde{X}|\tilde{Y}_0)).
\end{align}

\paragraph*{Alphabet Extension}

Now, let us consider the BSC with crossover probability $0<q<1$.
Since the BSC itself is not a GEC, Protocol \ref{protocol-GEC} cannot be used directly.
In order to apply Protocol \ref{protocol-GEC} to the BSC, the procedures of alphabet extension and erasure emulation
have been used in the literature \cite[Example 1]{AhlCsi13}. More specifically, the parties divide the index set $[n]$ into blocks of length $2$ 
(for simplicity, assume that $n$ is even number). For $i$th block, ${\cal P}_1$ randomly transmits
$(X_{i,1},X_{i,2})=(0,0)$ or $(X_{i,1},X_{i,2})=(1,1)$. Then, by setting ${\cal Y}_0=\{(0,0), (1,1)\}$ 
and ${\cal Y}_1 = \{(0,1), (1,0)\}$, this block channel can be regarded as a BSEC, an instance of GEC, such that erasure probability is $p_1=2q(1-q)$,
$W_0$ is equivalent to BSC of crossover probability $q_1=\frac{q^2}{(1-q)^2+q^2}$, and $W_1$ is equivalent to completely useless channel.
Thus, we can apply Protocol \ref{protocol-GEC} to this emulated BSEC and attain the lower bound
\begin{align} \label{eq:lower-bound-alphabet-extension}
C_{\mathtt{OT}}(W_{\rom{BSC}(q)}) \ge \frac{p_1}{2}(1-h(q_1)).
\end{align}

In \cite{SudWatYam:24}, we proposed a method to further improve the lower bound in \eqref{eq:lower-bound-alphabet-extension}.
Since the erasure probability of the above emulated BSEC is $p_1 < \frac{1}{2}$, blocks with fraction of $(1-2p_1)$ are discarded 
in Step \ref{protocol-GEC-step2} of Protocol \ref{protocol-GEC} (labeled as $V_i=2$ and not used in Steps \ref{protocol-GEC-step3}-\ref{protocol-GEC-step6}).
By reusing such discarded blocks to recursively emulate BSEC in $T$ rounds, we have shown that the following lower bound can be attained:
\begin{align} \label{eq:lower-bound-existing}
C_{\mathtt{OT}}(W_{\rom{BSC}(q)}) \ge \sum_{t=1}^T \bigg\{ \prod_{j=1}^{t-1} \frac{1 - 2 p_j}{2} \bigg\} \frac{p_t}{2}(1-h(q_t)),
\end{align} 
where $q_0:=q$ and
\begin{align}
q_t &:= 
\frac{q_{t-1}^2}{(1-q_{t-1})^2+q_{t-1}^2},  \label{eq:qt-existing} \\
p_t &:= 2 q_{t-1}(1-q_{t-1}). 
\end{align}
for $1 \le t \le T$.

\section{Improvement by Polarization} \label{section:polarization}

In the existing protocols, we emulated a BSEC from a BSC.
Our first idea of improvement is to emulate more general GECs that are not necessarily a BSEC.
To that end, for an integer $s$ and $N=2^s$, let us consider the generator matrix of polar code:
\begin{align*}
G_N = F^{\otimes s},~~~ F = \left(
\begin{array}{cc}
1 & 0 \\
1 & 1
\end{array}
\right),
\end{align*}
where $\otimes$ is the Kronecker product. We use $G_N$ to emulate GECs recursively. 

To fix an idea, let us consider the protocol for $N=2^2$.
The parties divide the index set $[n]$ into blocks of length $4$.
In the first round, for the first column $G_4(:,1)=(1,1,1,1)^\san{T}$ of $G_4$, let\footnote{All arithmetics are computed modulo $2$.} 
\begin{align*}
\tilde{{\cal X}}_1 := \{ \tilde{x} \in \{0,1\}^4: \tilde{x} \cdot G_4(:,1) = 0\}.
\end{align*}
For each block, ${\cal P}_1$ sends $\tilde{X}_1$ uniformly distributed on $\tilde{{\cal X}}_1$ over the BSC, and
${\cal P}_2$ receives $\tilde{Y}_1 \in \{0,1\}^4$. Then, by setting $\tilde{{\cal Y}}_{1,0} := \tilde{{\cal X}}_1$ as non-erasure set 
and $\tilde{{\cal Y}}_{1,1} := \{0,1\}^4\backslash \tilde{{\cal X}}_1$ as erasure set, the channel $P_{\tilde{Y}_1|\tilde{X}_1}$ can be regarded
as a GEC with erasure probability
\begin{align*}
p_1 := \sum_{\tilde{y} \in \tilde{{\cal Y}}_{1,1}} P_{\tilde{Y}_1|\tilde{X}_1}(\tilde{y}|\vec{0}) = 4q(1-q)((1-q)^2+q^2),
\end{align*} 
where $\vec{0}=(0,0,0,0)$;\footnote{Since the channel is additive, the erasure probability does not depend on the input sequence.}
and $P_{\tilde{Y}_1|\tilde{X}_1}$ can be decomposed as
\begin{align} \label{eq:decomposition-GEC-first-round}
P_{\tilde{Y}_1|\tilde{X}_1} = (1-p_1) P_{\tilde{Y}_{1,0}|\tilde{X}_1} + p_1 P_{\tilde{Y}_{1,1}|\tilde{X}_1},
\end{align}
where $\tilde{Y}_{1,b}$ is the random variable defined as $\tilde{Y}_1$ conditioned on $\tilde{Y}_1 \in \tilde{{\cal Y}}_{1,b}$.
In the first round, we apply Protocol \ref{protocol-GEC} to this emulated GEC $P_{\tilde{Y}_1|\tilde{X}_1}$.

Since there are discarded blocks with fraction of $(1-2p_1)$ in the first round, we proceed to the second round
for those blocks; note that, for those blocks, the receiver's observation is $\tilde{Y}_{1,0}$. 
For the second columns $G_4(:,2) = (0,1,0,1)^\san{T}$ of $G_4$, let 
\begin{align*}
\tilde{{\cal X}}_2 := \{ \tilde{x} \in \tilde{{\cal X}}_1: \tilde{x}\cdot G_4(:,2)=0\}.
\end{align*}
For each block, ${\cal P}_1$ reveals $\tilde{X}_1\cdot G_4(:,2)$ to ${\cal P}_2$. If the revealed value is $0$, then
the parties set $\tilde{X}_2 := \tilde{X}_1$ and $\tilde{Y}_2 := \tilde{Y}_{1,0}$, respectively; if the revealed value is $1$, then
the parties set $\tilde{X}_2:= \tilde{X}_1 + (0,0,1,1)$ and $\tilde{Y}_2 := \tilde{Y}_{1,0} + (0,0,1,1)$, respectively.\footnote{Instead of 
$(0,0,1,1)$, any vector in $\tilde{{\cal Y}}_{2,1}$ may be used.}
In this manner, by setting $\tilde{{\cal Y}}_{2,0} := \tilde{{\cal X}}_2$ as non-erasure set
and $\tilde{{\cal Y}}_{2,1}:= \tilde{{\cal X}}_1 \backslash \tilde{{\cal X}}_2$ as erasure set, we obtain a GEC with erasure probability
\begin{align*}
p_2 := \sum_{\tilde{y} \in \tilde{{\cal Y}}_{2,1}} P_{\tilde{Y}_2|\tilde{X}_2}(\tilde{y}|\vec{0}) = \frac{4q^2(1-q)^2}{1-p_1},
\end{align*}
and $P_{\tilde{Y}_2|\tilde{X}_2}$ can be decomposed as in \eqref{eq:decomposition-GEC-first-round}
for $P_{\tilde{Y}_{2,0}|\tilde{X}_2}$ and $P_{\tilde{Y}_{2,1}|\tilde{X}_2}$, where 
$\tilde{Y}_{2,b}$ is the random variable defined as $\tilde{Y}_2$ conditioned on $\tilde{Y}_2 \in \tilde{{\cal Y}}_{2,b}$.
In the second round, we apply Protocol \ref{protocol-GEC} to $P_{\tilde{Y}_2|\tilde{X}_2}$.

Again, for the discarded blocks with fraction $(1-2p_2)$, we proceed to the third round.
For the third column $G_4(:,3)=(0,0,1,1)^\san{T}$ of $G_4$, we can define $\tilde{X}_3$, $\tilde{Y}_3$, $\tilde{Y}_{3,0}$, $\tilde{Y}_{3,1}$,
$\tilde{{\cal X}}_3$, $\tilde{{\cal Y}}_{3,0}$ and $\tilde{{\cal Y}}_{3,1}$
in the same manner as the second round. Then, we obtain a GEC $P_{\tilde{Y}_3|\tilde{X}_3}$ with erasure probability
\begin{align*}
p_3 := \sum_{\tilde{y} \in \tilde{{\cal Y}}_{3,1}} P_{\tilde{Y}_3|\tilde{X}_3}(\tilde{y}|\vec{0}) = \frac{2q^2 (1-q)^2}{(1-p_1)(1-p_2)},
\end{align*}
and we can apply Protocol \ref{protocol-GEC} to $P_{\tilde{Y}_3|\tilde{X}_3}$.

By adding up the rates of the three rounds, we obtain a lower bound on the OT capacity:
\begin{align} \label{eq:polarize-N-4}
\sum_{t=1}^3 \bigg\{ \prod_{j=1}^{t-1}(1-2p_j) \bigg\} \frac{p_t}{4}\big( H(\tilde{X}_t|\tilde{Y}_{t,1}) - H(\tilde{X}_t|\tilde{Y}_{t,0}) \big).
\end{align}

For general $N=2^s$, our protocol proceeds as follows.
For each $t \in [N-1]$ and $t$th column $G_N(:,t)$ of $G_N$, we set
\begin{align*}
\tilde{{\cal X}}_t &:= \{ \tilde{x} \in \tilde{{\cal X}}_{t-1} : \tilde{x} \cdot G_N(:,t) = 0 \}, \\
\tilde{{\cal Y}}_{t,0} &:= \tilde{{\cal X}}_t, ~\tilde{{\cal Y}}_{t,1} := \tilde{{\cal X}}_{t-1}\backslash \tilde{{\cal X}}_t,
\end{align*}
where $\tilde{{\cal X}}_0 := \{0,1\}^N$. In the first round, ${\cal P}_1$ sends $\tilde{X}_1$ uniformly distributed on $\tilde{{\cal X}}_1$ over the BSC,
and ${\cal P}_2$ receives $\tilde{Y}_1$. In $t$th round for $t\ge 2$, ${\cal P}_1$ reveals $\tilde{X}_{t-1}\cdot G_N(:,t)$ to ${\cal P}_2$;
if the value is $0$, the parties set $\tilde{X}_t = \tilde{X}_{t-1}$ and $\tilde{Y}_t = \tilde{Y}_{t-1,0}$;  otherwise, the parties set
$\tilde{X}_t = \tilde{X}_{t-1}+\tilde{y}_{t,1}^{\min}$ and $\tilde{Y}_t= \tilde{Y}_{t-1,0}+\tilde{y}_{t,1}^{\min}$, where 
$\tilde{y}_{t,1}^{\min} \in \tilde{{\cal Y}}_{t,1}$ is the vector that is minimum in the lexicographic order,
and $\tilde{Y}_{t-1,b}$ is defined as $\tilde{Y}_{t-1}$ conditioned on $\tilde{Y}_{t-1} \in \tilde{{\cal Y}}_{t-1,b}$.
In this manner, we obtain a GEC $P_{\tilde{Y}_t|\tilde{X}_t}$ with erasure probability
\begin{align*}
p_t:= \sum_{\tilde{y}\in \tilde{{\cal Y}}_{t,1}} P_{\tilde{Y}_t|\tilde{X}_t}(\tilde{y}|\vec{0}).
\end{align*}
We apply Protocol \ref{protocol-GEC} to $P_{\tilde{Y}_t|\tilde{X}_t}$ recursively as long as $p_t\le \frac{1}{2}$.\footnote{
It can be verified that $p_1\le \frac{1}{2}$ for any $N=2^s$; for $N=4$, it can be verified that
$p_t\le \frac{1}{2}$ for every $t=1,2,3$; for $N=8$ and $16$, we numerically verified that $p_t\le \frac{1}{2}$ for every $t \in [N-1]$.}
In total, we obtain the following lower bound on the OT capacity:
\begin{align} \label{eq:lower-bound-polarization}
\sum_{t=1}^{L} \bigg\{ \prod_{j=1}^{t-1} (1-2p_j) \bigg\} \frac{p_t}{N} \big( H(\tilde{X}_t|\tilde{Y}_{t,1}) - H(\tilde{X}_t|\tilde{Y}_{t,0}) \big),
\end{align}
where $L$ is the maximum integer such that $p_L\le \frac{1}{2}$.

\begin{figure}[tb]
\centering{
\includegraphics[width=0.9\linewidth]{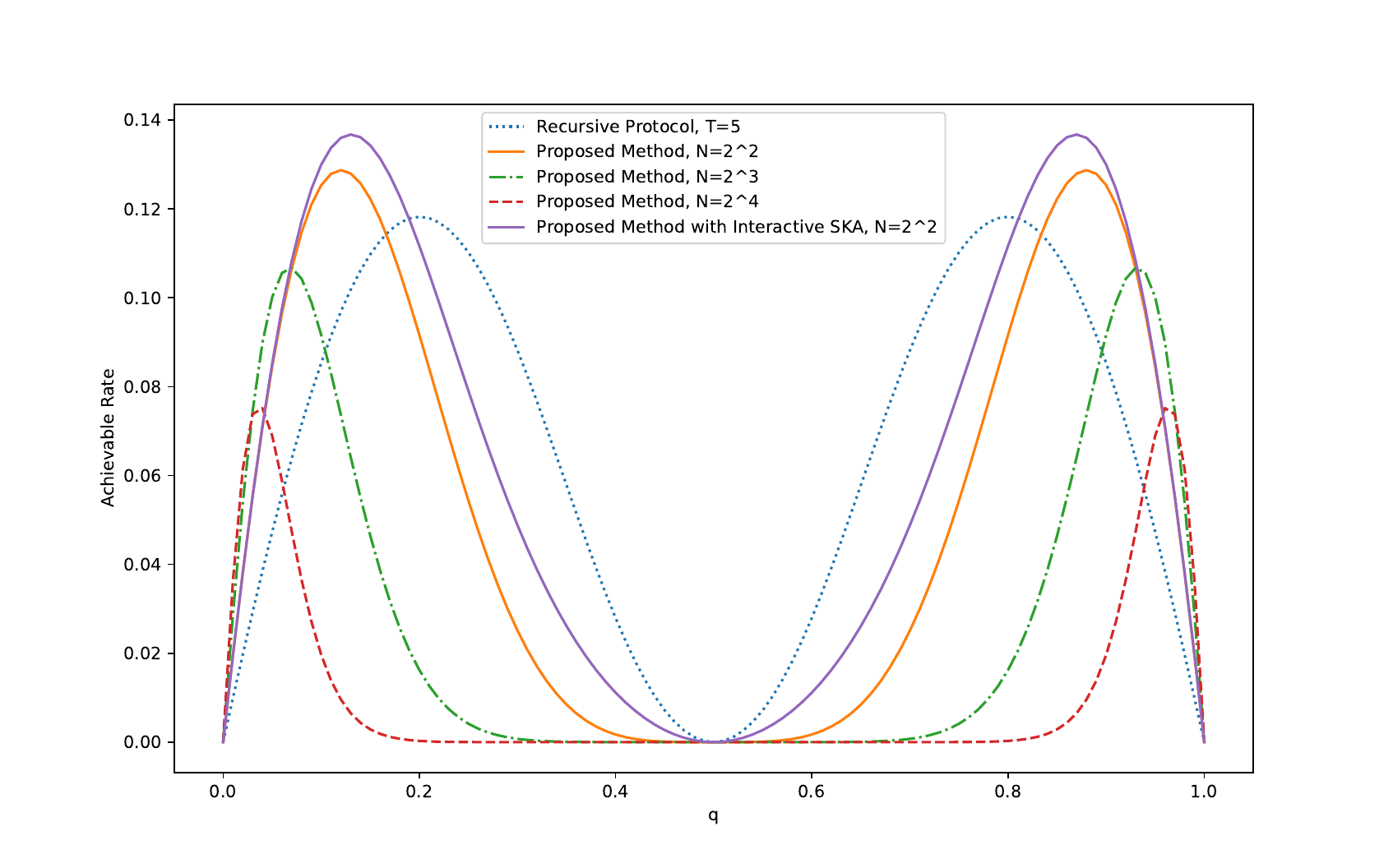}
\caption{A comparison of the existing lower bound \eqref{eq:lower-bound-existing} on the OT capacity 
for $T=5$ (``Recursive Protocol $T=5$"), the improved lower bound on the OT capacity in \eqref{eq:lower-bound-polarization}
for $N=4,8,16$ (``Proposed Method"),
and the further improved lower bound with interactive key agreement in Section \ref{section:interactive-key}
(``Proposed Method with Interactive SKA"), 
where the horizontal axis is the crossover probability $0 \le q \le 1$ of BSC.}
\label{Fig:comparison}
}
\end{figure}

In Fig.~\ref{Fig:comparison}, we plot the lower bound \eqref{eq:lower-bound-polarization} for $N=2^2, 2^3, 2^4$
and the existing lower bound \eqref{eq:lower-bound-existing} for $T=5$.\footnote{The lower bound \eqref{eq:lower-bound-existing} barely changes for larger $T$.}
As we can see from the figure, the lower bound by the proposed method is superior to the existing one when the crossover probability $q$
is rather small (slightly below $q=0.2$) or rather large (slightly above $q=0.8$). Furthermore, as $N$ increases, we can see that 
the slope of tangent at $q=0$ becomes larger. In order to further investigate this behavior, we derive more explicit form 
of the first term of the lower bound \eqref{eq:lower-bound-polarization}.
We derive the formula by noting a recursive structure
of the polarization matrix $G_N$; a proof will be provided in Appendix \ref{appendix:proof-recursive-formula}.

\begin{proposition} \label{proposition:recursive-formula}
Let $\bar{p}_0:=q$, and let
\begin{align*}
\bar{p}_j := \frac{1-(1-2q)^{2^j}}{2},~\bar{q}_j := \frac{\bar{p}_{j-1}^2}{1-\bar{p}_j}
\end{align*}
for $1 \le j \le s$.
Then, for $N=2^s$, we have
\begin{align*}
\lefteqn{ \frac{p_1}{N}\big(H(\tilde{X}_1|\tilde{Y}_{1,1}) - H(\tilde{X}_1|\tilde{Y}_{1,0}) \big) } \\
&= \frac{\bar{p}_s}{N}\sum_{i=0}^{s-1}\bigg\{
\prod_{j=1}^i (1-2\bar{q}_{s-j+1}) \bigg\}
\big(1 - h(\bar{q}_{s-i})\big) =: f(q).
\end{align*}
Furthermore, the derivative of $f(q)$ at $q=0$ satisfies 
\begin{align} \label{eq:derivative-of-odd-even}
\lim_{q\downarrow 0} f^\prime(q)= s.
\end{align}
\end{proposition}

From \eqref{eq:derivative-of-odd-even} of Proposition \ref{proposition:recursive-formula},
we can see that the slope of tangent of the lower bound at $q=0$ becomes arbitrarily large 
as $N=2^s$ increases. On the other hand, it has been known that the OT capacity of BSC is upper bounded 
by $h(q)$ \cite{AhlCsi13}; note that the slope of tangent of the upper bound at $q=0$ is unbounded as well.
Thus, the tangential behavior of the upper and lower bounds are now matching; previously,
it was not known whether the slope of the tangent of the OT capacity at $q=0$
is unbounded or not.

\section{Improvement by Interactive Key Agreement} \label{section:interactive-key}

In this section, we present another method to further improve the lower bound in Section \ref{section:polarization}.
We use the same notations as in Section \ref{section:polarization}, but we focus on $N=4$.
The main idea is to use the interactive key agreement in a manner oblivious to ${\cal P}_1$.
For an exposition of the interactive key agreement, see \cite[Sec.~10.5]{TyaWat:book}.

When we apply Protocol \ref{protocol-GEC} to $P_{\tilde{Y}_1|\tilde{X}_1}$, ${\cal P}_1$ sends the hash of
observations at rate $H(\tilde{X}_1|\tilde{Y}_{1,0})$ so that ${\cal P}_2$ can reproduce $\tilde{X}_1$ for
indices in ${\cal I}_{B}$. In the interactive key agreement, instead of $\tilde{X}_1$ itself, ${\cal P}_2$ seeks to
reproduce partial information of $\tilde{X}_1$ in multiple rounds by interactively communicating with ${\cal P}_1$.

Let 
\begin{align*}
U_1 &:= (\tilde{X}_1 \cdot G_4(:,2), \tilde{X}_1 \cdot G_4(:, 3)), \\
U_2 &:= \bol{1}[ (\tilde{Y}_{1,0} \cdot G_4(:,2), \tilde{Y}_{1,0}\cdot G_4(:,3)) = U_1], \\
U_3 &:= \left\{
\begin{array}{ll}
\tilde{X}_1 \cdot G_4(:,4) & \mbox{if } U_2=1 \\
\bot & \mbox{if } U_2 = 0
\end{array}
\right.,
\end{align*}
where $\bot$ indicates that it is discarded and set to be constant.
A rational of discarding $\tilde{X}_1$ when $U_2=0$ is that the remaining uncertainty is $H(\tilde{X}_1|\tilde{Y}_{1,0},U_1,U_2=0)=1$
and $\tilde{X}_1$ is useless for creating secret key.
In our interactive key agreement procedure, 
${\cal P}_1$ first sends the hash of $U_1$ at rate $H(U_1|\tilde{Y}_{1,0})$ so that ${\cal P}_2$ can reproduce $U_1$ for
indices in ${\cal I}_B$; for those indices, ${\cal P}_2$ computes $U_2$, and sends $U_2$ back to ${\cal P}_1$.
Then, ${\cal P}_1$ computes $U_3$ and sends the hash of $U_3$ at rate $H(U_3|\tilde{Y}_{1,0},U_2)$
so that ${\cal P}_2$ can reproduce $U_3$.

If this interactive key agreement procedure is used just for key agreement against ``external adversary," then there is no problem.
However, in the OT protocol, ${\cal P}_2$'s input $B$ must be concealed from ${\cal P}_1$. If ${\cal P}_2$ sends $U_2$ back to ${\cal P}_1$
only  for indices in ${\cal I}_B$, then ${\cal P}_1$ can recognize those indices are good indices and know the value of $B$.
To avoid that, ${\cal P}_2$ must behave the same for indices in both ${\cal I}_0$ and ${\cal I}_1$.
Fortunately, $U_2$ only depends on the difference between $\tilde{X}_1$ and $\tilde{Y}_{1,0}$, and thus
$U_2$ is independent of $\tilde{X}_1$. Thus, for indices in ${\cal I}_{\overline{B}}$, ${\cal P}_2$ samples 
$U_2^\prime$ according to the same marginal distribution of $U_2$, and sends it to ${\cal P}_1$.
Then, ${\cal P}_1$ computes (without recognizing that the index is in ${\cal I}_{\overline{B}}$)
\begin{align*}
U_3^\prime &:= \left\{
\begin{array}{ll}
\tilde{X}_1 \cdot G_4(:,4) & \mbox{if } U_2^\prime = 1 \\
\bot & \mbox{if } \tilde{U}_2^\prime = 0
\end{array}
\right..
\end{align*}

For indices in ${\cal I}_{\overline{B}}$, ${\cal P}_2$'s ambiguity about $(U_1,U_3^\prime)$ is $H(U_1,U_3^\prime|Y_{1,0},U_2^\prime)$; 
in the information reconciliation, ${\cal P}_1$ sends hashes at rates $H(U_1|\tilde{Y}_{1,0})$ and $H(U_3|\tilde{Y}_{1,0},U_2)$.
Thus, from the results on the information reconciliation and the privacy amplification, the $t=1$ term of the lower bound in \eqref{eq:polarize-N-4}
can be improved to
\begin{align*}
\frac{p_1}{4}\big( H(U_1,U_3^\prime|Y_{1,0},U_2^\prime) - H(U_1|\tilde{Y}_{1,0}) -  H(U_3|\tilde{Y}_{1,0},U_2) \big).
\end{align*} 
The improved lower bound is also plotted in Fig.~\ref{Fig:comparison}.



\section{Discussions} \label{section:discussion}

In this paper, we proposed two new methods for the OT construction; by using these methods,
we derived lower bounds that are superior to the existing one when the crossover probability $q$ is rather 
small or rather large. Even though the proposed methods are not effective when $q$ is close to $\frac{1}{2}$,
since $q_t$ in \eqref{eq:qt-existing} satisfies $q_{t+1}<q_t$, i.e., strictly decreasing by iteration, for $q<\frac{1}{2}$,
it is possible to combine the proposed methods after appropriate number of iterations of the protocol in \cite{SudWatYam:24} to 
slightly improve \eqref{eq:lower-bound-existing} for $q$ close to $\frac{1}{2}$. 

Apart from the tangential behavior at $q=0$, there is still a gap between the upper bound and lower bounds.
In order to close the gap, we need to introduce more systematic methods to derive lower bounds;
also, at this point, there is no clue whether the upper bound in \cite{RaoPra14} is tight or not.


\section*{Acknowledgment}

This work was supported in part by the Japan Society for the Promotion of Science (JSPS)
KAKENHI under Grant 23H00468 and 23K17455.

\newpage
\pagebreak

\appendix
\subsection{Proof of Proposition \ref{proposition:recursive-formula}} \label{appendix:proof-recursive-formula}

For $1 \le j \le s$, let $E_j$ be i.i.d. Bernoulli random variable of length $2^j$ such that
the probability of each coordinate being $1$ is $q$. 
For $1 \le j \le s$ and $b=0,1$, let 
\begin{align*}
\tilde{{\cal Y}}_b^j := \{ \tilde{y} \in \{0,1\}^{2^j} : \tilde{y} \cdot G_{2^j}(:,1) = b\}.
\end{align*}
Then, note that $H(\tilde{X}_1|\tilde{Y}_{1,b}) = H(E_s|E_s \in {\cal Y}_1^s)$.
To compute this conditional entropy recursively, observe that the set $\tilde{{\cal Y}}_b^j$ has the following recursive structure:
\begin{align*}
\tilde{{\cal Y}}_0^j &= \big( \tilde{{\cal Y}}_0^{j-1} \times \tilde{{\cal Y}}_0^{j-1} \big) \cup \big( \tilde{{\cal Y}}_1^{j-1}\times \tilde{{\cal Y}}_1^{j-1} \big),\\
\tilde{{\cal Y}}_1^j &= \big( \tilde{{\cal Y}}_0^{j-1} \times \tilde{{\cal Y}}_1^{j-1} \big) \cup \big( \tilde{{\cal Y}}_1^{j-1} \times \tilde{{\cal Y}}_0^{j-1} \big).
\end{align*}
Thus, we have
\begin{align*}
\Pr( E_j \in \tilde{{\cal Y}}_1^j) = 2 \cdot \Pr( E_{j-1} \in \tilde{{\cal Y}}_0^{j-1}) \cdot \Pr( E_{j-1} \in \tilde{{\cal Y}}_1^{j-1}). 
\end{align*}
By noting the recursive formula
\begin{align*}
\bar{p}_j &= \frac{1-(1-2q)^{2^j}}{2} \\
&= 2 \cdot \frac{1+(1-2q)^{2^{j-1}}}{2} \cdot \frac{1-(1-2q)^{2^{j-1}}}{2} \\
&= 2 (1-\bar{p}_{j-1})\bar{p}_{j-1},
\end{align*}
we can recursively prove
\begin{align*}
\Pr( E_j \in \tilde{{\cal Y}}_1^j ) = \bar{p}_j.
\end{align*}
Furthermore, by noting 
\begin{align*}
\Pr( E_j \in \tilde{{\cal Y}}_1^{j-1}\times \tilde{{\cal Y}}_1^{j-1} | E_j \in \tilde{{\cal Y}}_0^j) &= \bar{q}_j, \\
\Pr( E_j \in \tilde{{\cal Y}}_1^{j-1} \times \tilde{{\cal Y}}_0^{j-1} | E_j \in \tilde{{\cal Y}}_1^j) &= \frac{1}{2},
\end{align*}
we have the following recursive formulae:
\begin{align*}
H(E_j | E_j \in \tilde{{\cal Y}}_0^j) &= h(\bar{q}_j) \\
&~~~+ (1- \bar{q}_j) \cdot 2 \cdot H(E_{j-1}|E_{j-1} \in \tilde{{\cal Y}}_0^{j-1}) \\
&~~~+ \bar{q}_j \cdot 2 \cdot H(E_{j-1} | E_{j-1} \in \tilde{{\cal Y}}_1^{j-1}), \\
H(E_j|E_j \in \tilde{{\cal Y}}_1^j) &= 1 + \big( H(E_{j-1}|E_{j-1} \in \tilde{{\cal Y}}_0^{j-1}) \\
&~~~+ H(E_{j-1}|E_{j-1}\in \tilde{{\cal Y}}_1^{j-1}) \big)
\end{align*}
and
\begin{align*}
\lefteqn{ H(E_j|E_j \in \tilde{{\cal Y}}_1^j)  - H(E_j | E_j \in \tilde{{\cal Y}}_0^j) } \\
&= 1 - h(\bar{q}_j)
+ (1-2 \bar{q}_j) \big( H(E_{j-1}|E_{j-1}\in \tilde{{\cal Y}}_1^{j-1}) \\
&\hspace{40mm} - H(E_{j-1}|E_{j-1} \in \tilde{{\cal Y}}_0^{j-1}) \big).
\end{align*}
By solving this recursive formula, we have
\begin{align*}
\lefteqn{ H(E_s|E_s \in \tilde{{\cal Y}}_1^s)  - H(E_s | E_s \in \tilde{{\cal Y}}_0^s) } \\
&= \sum_{i=0}^{s-1} \bigg\{ \prod_{j=1}^i (1- 2 \bar{q}_{s-j+1}) \bigg\} \big(1 - h(\bar{q}_{s-i})\big).
\end{align*}
By combining this with $p_1 = \bar{p}_s$, we have the first claim of the proposition.

To obtain the derivative, note that
\begin{align*}
\bar{p}_j^\prime &= 2^j(1-2q)^{2^j-1}, \\
\bar{q}_j^\prime &= \frac{2 \bar{p}_{j-1} \bar{p}_{j-1}^\prime (1-\bar{p}_j)+\bar{p}_{j-1}^2 \bar{p}_j^\prime}{(1-\bar{p}_j)^2}, \\
h^\prime(q) &= \log \frac{1-q}{q}.
\end{align*}
Note also that $\bar{p}_j\big|_{q=0}=0$, $\bar{q}_j \big|_{q=0}=0$, 
and $\bar{q}_j^\prime \big|_{q=0}=0$. In fact, the numerator of $\bar{q}_j^\prime$ is a polynomial of $q$.
Thus, we have
\begin{align*}
\lim_{q\downarrow 0} \frac{d}{dq}h(\bar{q}_j^\prime) = \lim_{q\downarrow 0} \bigg( \log \frac{1-\bar{q}_j}{\bar{q}_j} \bigg)\cdot \bar{q}_j^\prime = 0.
\end{align*} 
By computing the derivative of $f(q)$ via the product rule and setting $q\downarrow 0$, all terms other than the first vanish and we obtain
\begin{align*}
\lim_{q\downarrow 0} f^\prime(q) = 
\lim_{q\downarrow 0} \frac{\bar{p}_s^\prime}{N} \sum_{i=0}^{s-1}\bigg\{ \prod_{j=1}^i (1-2 \bar{q}_{s-j+1})  \bigg\}\big( 1 - h(\bar{q}_{s-i}) \big).
\end{align*}
By substituting $\bar{p}_s^\prime = 2^s(1-2q)^{2^s-1}$ computed above 
and there are $s$ terms of $1$'s in the second factor, we have \eqref{eq:derivative-of-odd-even}.
\qed


\bibliographystyle{../../../09-04-17-bibtex/IEEEtran}
\bibliography{../../../09-04-17-bibtex/reference.bib}
\end{document}